\documentclass{PoS}

\title{Fermi-LAT constraints on diffuse Dark Matter annihilation from the Galactic Halo }

\ShortTitle{Fermi-LAT constraints on Galactic Dark Matter annihilation}

\author{\speaker{Brandon ANDERSON}%
        \thanks{On behalf of the Fermi-LAT Collaboration.}\\
       University of California - Santa Cruz\\
       E-mail: \email{anderson@physics.ucsc.edu}}

\abstract{Our Galaxy resides in the center of a vast "Halo" of Dark Matter (DM). 
This concentration produces, in many viable particle physics models, an indirect Weakly 
Interacting Massive Particle (WIMP) annihilation signal that peaks in the Fermi-LAT's energy 
range. Our knowledge of the diffuse background is essential to placing reasonable limits on 
the DM mass and cross-section.  We incorporate a systematic variation of the GALPROP galactic 
diffuse background model, constrained by current cosmic-ray measurements, into a profile likelihood 
analysis and present preliminary upper limits on the DM annihilation cross-section using the Fermi-LAT data.}

\FullConference{Identification of Dark Matter 2010-IDM2010\\
		July 26-30, 2010\\
		Montpellier France}

\begin{document}

\section{Signal}

The Milky Way rests at the center of a massive, $\mathcal{O}(10^{12})~M_{\odot}$ \cite{Diemand2007}, 
Halo of Dark Matter (DM).  We have a good sense of the large-scale mass distribution of this Halo from N-body simulations, 
and can approximate that with a simple density 
profile function.  For example, here we use an Einasto profile, where $\rho(r) \propto \exp{(-Ar^\alpha)}$.  If this Halo 
is comprised of Weakly Interacting Massive Particles (WIMPs) capable of annihilation or decay into 
Standard Model (SM) particles, the Fermi-LAT provides a unique all-sky energy window to search for gamma-ray 
signal resulting from that process.

To capture the uncertainty in our DM signal, we consider WIMPs of mass 25 to 500 GeV annihilating to a variety 
of SM channels, including $b\bar{b}$, $t\bar{t}$, $\tau^{+}\tau^{-}$, and $\mu^{+}\mu^{-}$.  And while N-body 
simulations are powerful, their inability to resolve small scale structure leaves another uncertainty in the ``boost''
that comes from extrapolating their mass functions down to the free-streaming cutoff.  We therefore use three 
different mass function exponents (1.8,1.9,2.0) to cover the reasonable \cite{Diemand2008} range of possibilities.

While channels like $b\bar{b}$ deliver their gamma-ray signal mostly through $\pi^{0}$ decay, others
(e.g. $\mu^{+}\mu^{-}$) require an extra step in calculating the predicted Fermi-LAT signal.  This is the 
propagation and interaction of the final SM decay products.  
This calculation is performed with the same methodology as the rest of our cosmic-ray (CR) backgrounds discussed 
in the next section.

\section{Diffuse Backgrounds}

Since we expect the Halo signal to be large-scale and slowly varying, masking out regions of the sky containing 
point sources greatly simplifies our analysis without much affecting our sample size.  We mask 
the entire galactic plane to $\pm10^{\circ}$, and Fermi sources with an energy-dependent sized mask 
corresponding to the LAT PSF.  

After masking, we contend only with diffuse backgrounds to the Halo signal: extragalactic, 
instrumental, point source residuals, and galactic diffuse.  The first two enter the fit as completely isotropic 
signals with fixed energy spectra \cite{extragalactic}.  Point source residuals, products of the extended 
tails of the PSF 
that reach outside our masking scheme, are modeled using Fermi-LAT software 
\footnote{\underline{http://Fermi.gsfc.nasa.gov/ssc}} tuned to the Fermi First Source Catalogue \cite{1fgl} measured 
source parameters.  

Originating from interactions (bremsstraulung, inverse compton, and $\pi_{0}$ decay) of CR 
with the gas, light, dust, and magnetic fields in the Milky Way, the galactic diffuse is by far the most difficult background 
to model.  GALPROP \cite{Strong2000}\ is a code designed to self-consistently solve the transportation and 
interaction of CR 
within the galaxy from source injection to arrival at Earth radius. We assess the quality of a particular 
realization of GALPROP's parameter space by comparing with available CR measurements.  This is 
advantageous in the respect that it makes our model independent of the gamma rays we wish to probe for DM signals.

The large number and uncertainty of parameters in the GALPROP framework, however, requires us to make a thorough 
investigation of that space to quantify our systematics.  To that end, we fit with models generated within the 
parameter space in Table~\ref{pspace}.  Each model's validity is quantified by a $\chi^{2}$ fit to ${}^{10}$Be/${}^{9}$Be,
B/C, and proton data from the HEAO-3, IMP, ATIC-2, CREAM, ACE, ISOMAX, AMS01, CAPRICE, and BESS experiments.  
The explicit form is, 

\begin{equation}
 \chi^2 = \Sigma_j \Sigma_i^{Nj} \frac{(D_{ij} - T_{ij})^2}{\sigma^2_{ij}+\Delta\phi^{2}_{ij}},
\label{chisqeq}
\end{equation}

where D and T are model and data respectively, $\Delta\phi^{2}$ comes from uncertainty in the solar modulation (taken to be 
$\pm100$MV), and the sum is over all points and experiments.  

One uncertainty obviously missing from the parameter scan is the normalization of the primary electron sources.
The sensitivity of a local measurement to nearby sources (or lack of) makes this a particularly difficult value 
to set a parameter range for.  Running each model with and without primary electrons and finding the difference isolates 
their contribution so that we can let it float freely in our fits to the data.

\begin{table}
\begin{center}
\begin{tabular}{ |l|l| p{0.2cm} | }
\hline
 \textbf{Parameter} & \textbf{Range} \\
\hline \hline
 Diffusion Coefficient & 1$\times10^{27} \rightarrow$ 4$\times10^{29}$ \\
 Halo Height & 1 $\rightarrow$ 11 kpc \\
 Diffusion Index & 0.33, 0.50\\
 Alfven Velocity & 0 $\rightarrow$ 50 km s$^{-1}$\\
 Electron Injection Index & 1.8 $\rightarrow$ 2.5\\
 Nucleon Injection Index (Low) & 1.7 $\rightarrow$ 2.6\\
 Nucleon Injection Index (High) & 2.26, 2.43\\
 Source Distribution & Parameterized, SNR, Pulsars\\
 \hline
\end{tabular}
\caption{GALPROP parameter space.}
\label{pspace}
\end{center}
\end{table}

\section{Profile Likelihood}


After binning into 12 annular and 80 logarithmic energy bins, we compare our model with the Fermi-LAT gamma-ray data and 
find the Poisson Likelihood, $L$ \cite{mattox1996}.  For each DM channel and mass we then produce a Maximum Likelihood,

\begin{equation}
 \hat{L}_{j}(\theta_{DM})=\prod_{i} P_{ij}(n_{i};\vec{\alpha}_{max},\theta_{DM}),
\label{leq}
\end{equation}

by selecting the best-fitting linear parameters, $\vec{\alpha}$ (EGB, instrumental, and primary electron normalizations),
for each DM normalization, $\theta_{DM}$.  The $j$ represents the diffuse model and the product is over each bin, $i$.

\begin{figure}
\begin{center}
\includegraphics[width=0.5\textwidth]{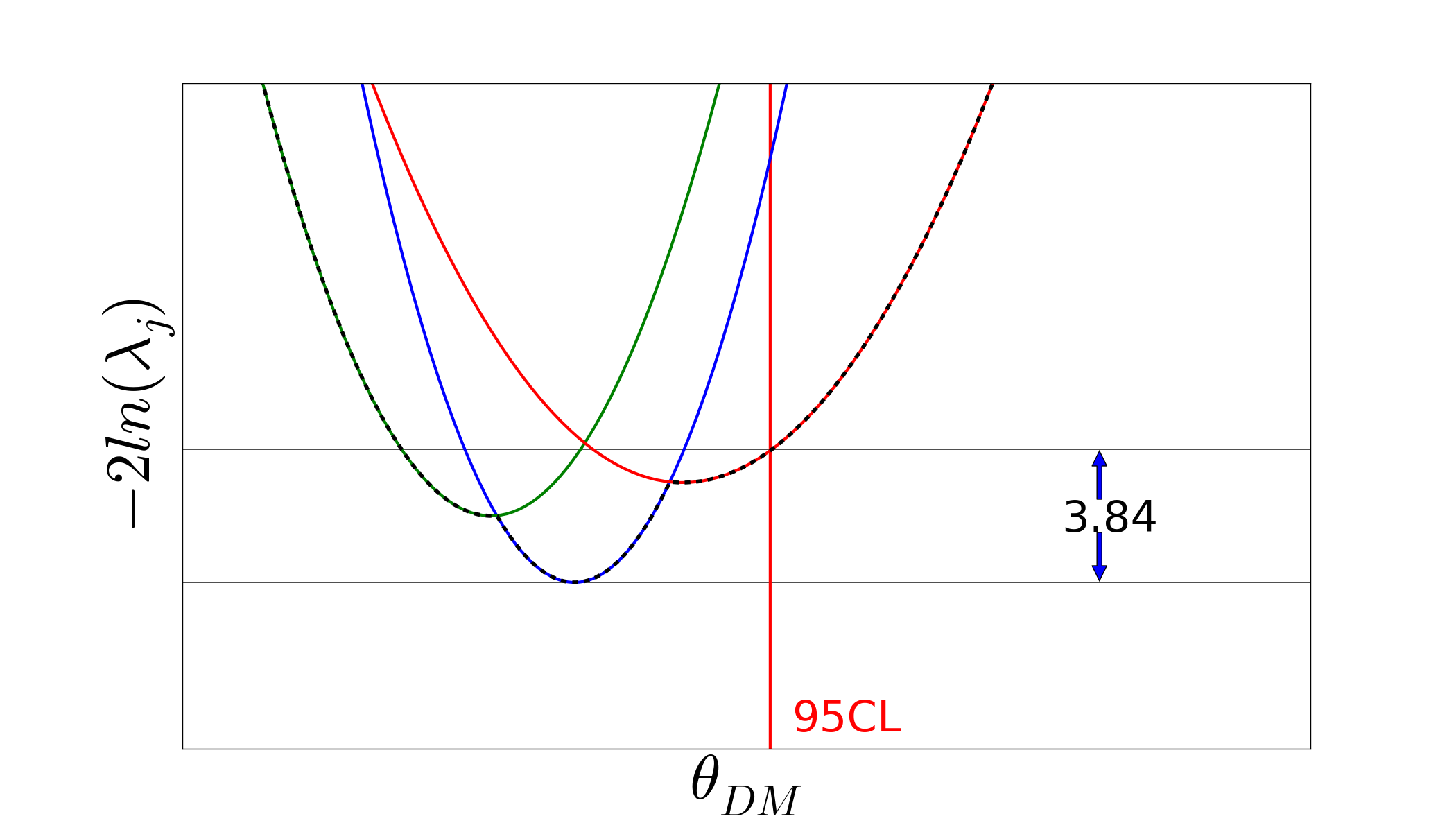}

\end{center}
\caption{Qualitative plot of an upper limit set using the Profile Likelihood method on the Test 
Statistic defined in Eq. \protect \ref{tseq}.  Each curve represents a different GALPROP model.}
\label{fig1}
\end{figure}

Finally, we convert $\hat{L}$ into a Test Statistic for every GALPROP model that includes 
the CR-$\chi^{2}$ (Eq.~\ref{chisqeq}) information,

\begin{equation}
 \lambda_{j}(\theta_{DM})=\frac{P_{j}^{CR}\hat{L}_{j}(\theta_{DM})}{(P_{j}^{CR}\hat{L}_{j})_{best}}.
\label{tseq}
\end{equation}

This leaves us with a family of curves (Fig.~\ref{fig1}), the minimum contour of which comprises the Profile Likelihood.  
The contour should behave as a $\chi^{2}$ with one degree of freedom, and we set our 95\% confidence upper 
limit on $\theta_{DM}$ to where it rises above the minimum by 3.84. 

\begin{figure}
\begin{center}
\includegraphics[width=.6\textwidth]{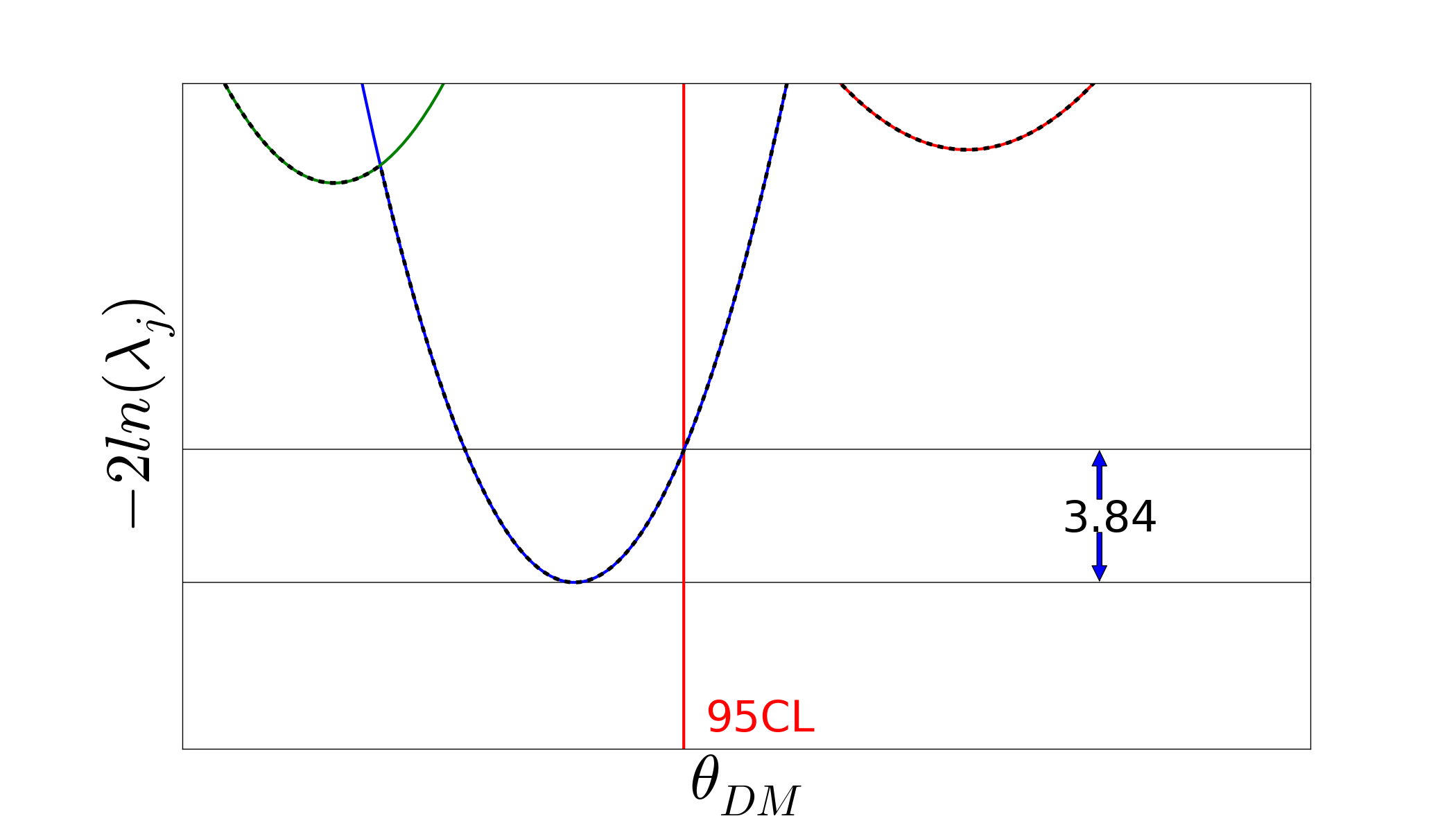}
\caption{A qualitative plot of an under-sampled minimum in the Profile Likelihood.}
\label{fig2}
\end{center}
\end{figure}

To be confident with the limit it is necessary to have sufficiently populated the model space such that the minimum 
is the \emph{true} minimum, and that the nearby sampling is dense enough to smoothly detect the profile out to the 
upper limit.  Under-sampling the parameter space leads to a situation like in Figure~\ref{fig2}, where the limit is derived 
from a single model, and both the location of the true minimum and the shape of the curve around it are unclear.  

\begin{figure}
\begin{center}
\includegraphics[width=.6\textwidth]{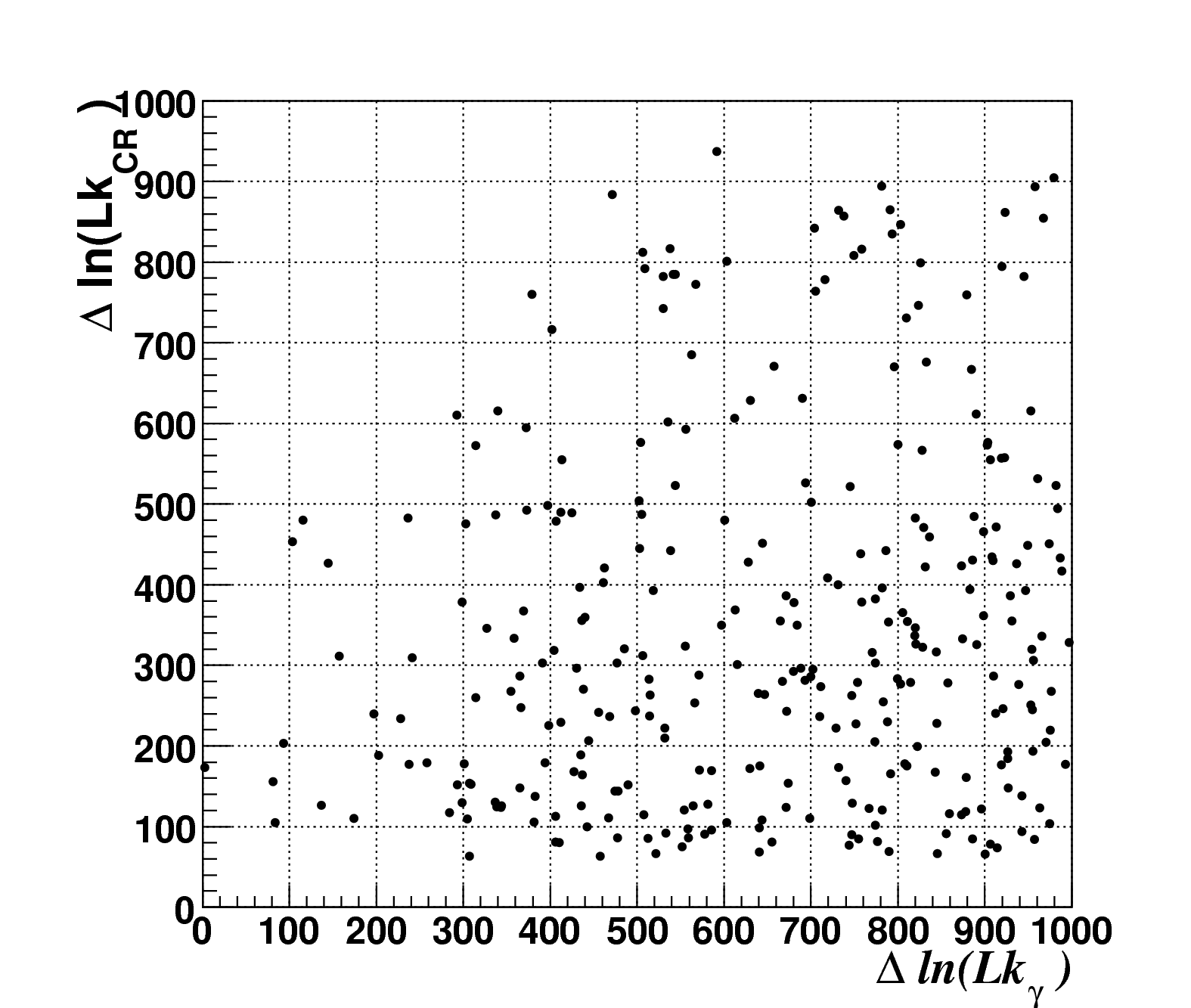}
\caption{A scatter plot of the CR vs. gamma ray maximum Likelihood for our present models.}
\label{crvsgamma}
\end{center}
\end{figure}

After our initial, completely random, sampling of the parameter space in Table~\ref{pspace}, we found ourselves in precisely 
that situation.  Figure~\ref{crvsgamma} illustrates the initial sampling as a scatter plot for the CR and gamma rays $\hat{L}$.
The minimum will be near the lower left hand corner, and sampling should be densest around it.  Clearly this was 
not the case.

\subsection{Classification Tree Sampling}

Being limited mainly by computation time, the best way to populate a specific region is to have a rough idea what its
$\hat{L}$ will be beforehand.  Predicting the gamma-ray fits would be problematic, due to the variety of DM 
signals being explored.  The CR-$\chi^{2}$ is much more straight forward, and so we use it to train a classification 
tree.  The tree can then be used to pre-select GALPROP models that agree with CR data.  This will reduce much of the y-axis 
scatter in Fig.\ref{crvsgamma}, hopefully allowing us to sufficiently sample the global minimum with our available 
computer resources.

\section{Preliminary Results}

\begin{figure}
\begin{center}
\includegraphics[width=.6\textwidth]{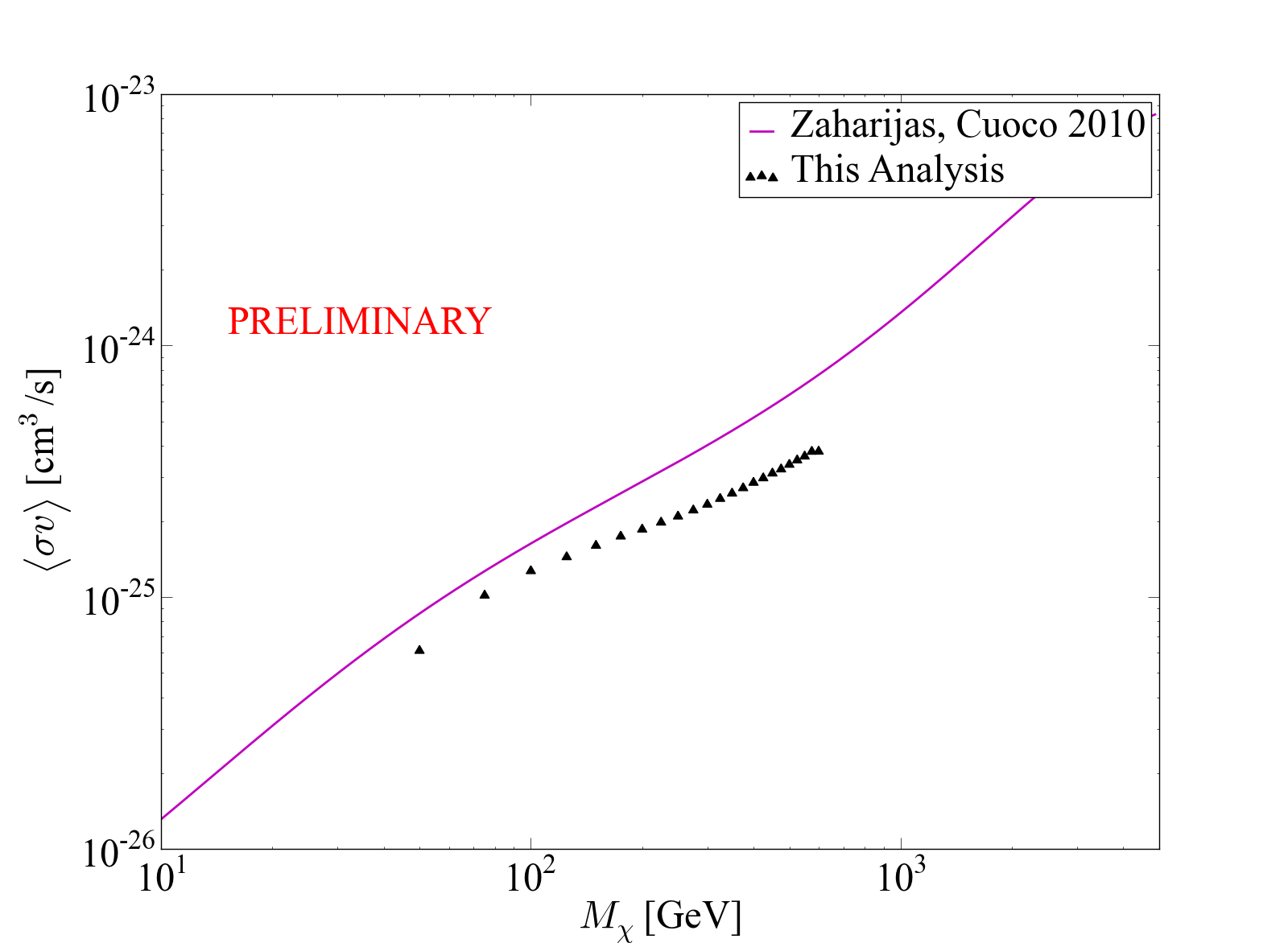}
\caption{Present 95\% upper limits on pure $b\bar{b}$ annihilation from boosted Einasto DM.  Also shown are $3\sigma$ 
limits from an independent Fermi Collaboration analysis with an unboosted NFW profile, 4 kpc Halo height, and SNR-based 
CR source distribution \protect \cite{gzah2010} \cite{Cuoco2010}.}
\label{prelimlim}
\end{center}
\end{figure}

Figure \ref{prelimlim} shows our present 95\% confidence upper limits on a pure $b\bar{b}$ channel WIMP.  As mentioned earlier, 
these limits are based on a profile likelihood minimum composed of a single GALPROP model.  This makes the limits 
very preliminary and it is unclear in which direction the final ones will move.  Also shown in Fig. \ref{prelimlim} 
are the closely-related limits derived by an independent Fermi Collaboration analysis which uses a single ``conservative''
GALPROP model with all of its gas (target) map contributions free to float.  

A further caveat, there are persistent residuals even in our best fits.  We hope to account for much of this by correcting 
known deficiencies to our model, foremost being the isotropic approximation to the inverse compton signal.  We also plan on
addressing uncertainty in the gas distributions by allowing some float in the collective contributions produced by
interaction with molecular, atomic, and ionized Hydrogen.  Given the level of uncertainties discussed above, these upper 
limits should be considered as purely illustrative of the Fermi LAT potential to constrain DM annihilation or decay from 
the Galactic halo.

\section{Summary and Outlook}

We have conducted a preliminary investigation of possible DM annihilation signal in the Milky Way Host Halo, and 
within the systematic uncertainties of our background model, find no significant detection.  We 
model all known backgrounds, and use the Profile Likelihood method of marginalizing over the many nuisance parameters 
that come from uncertainty in the CR-induced galactic diffuse emission.  Sampling the model parameter space adequately 
is a computationally difficult task, one we are addressing by intelligently pre-selecting our models with a classification
tree.  We expect the combination of increased sampling density of our maximum likelihood region and upcoming improvements
to our models (anisotropic inverse compton \& inclusion of target uncertainty) to yield a robust limit that takes into account the large systematics inherent to a 
physical model of the galactic background.


\begin{thebibliography}{99}

\bibitem[Abdo et al.~(2010)]{extragalactic}
Abdo, A.A. et al. \ 2010, \emph{Phys. Rev. Lett.} 104, 101101.

\bibitem[Abdo et al.~(2010)]{1fgl}
Abdo, A.A. et al. \ 2010, \emph{APJS} 188 405.

\bibitem[Cuoco et al.\ (2010)]{Cuoco2010}
Cuoco, A. et al. \ 2010, \emph{Identification of Dark Matter Conference}.

\bibitem[Diemand et al.\ (2007)]{Diemand2007}
Diemand, J., Kuhlen, M., Madau, P.\ 2007, \emph{APJ} 667, 859.

\bibitem[Diemand et al.\ (2008)]{Diemand2008}
Diemand, J., Kuhlen, M., Madau, P., Zemp, M., Moore, B., Potter, D., Stadel, J.\
2008, \emph{Nature}, 454, 735.

\bibitem[Mattox et al.~(1996)]{mattox1996}
Mattox, J. et al.\ 1996, \emph{APJ} 461, 396.

\bibitem[Strong et al.\ (2000)]{Strong2000}
Strong, A.~W., Moskalenko, I.~V., \& Reimer, O.\ 2004, \emph{APJ}, 537, 763.

\bibitem[Zaharijas et al.\ (2010)]{gzah2010}
Zaharijas, G. et al. \ 2010, \emph{Identification of Dark Matter Conference}.

\end{thebibliography}
\end{document}